\documentclass[12pt]{article}
\usepackage[dvips]{graphicx}
\begin{document}
\title{Quarkonium correlators at finite temperature and potential models}
\author{\'Agnes M\'ocsy\\
        RIKEN-BNL Research Center, Brookhaven National Laboratory, \\
        Upton,  NY 11973  USA,\\[3mm]
P\'eter Petreczky,\\
       Department of Physics, Brookhaven National Laboratory,\\
 Upton, NY 11973 USA \\
        }

\maketitle
\begin{abstract}
We discuss the calculations of quarkonium spectral functions in potential
models and their implications for the interpretation of the lattice data
on quarkonium correlators. In particular, we find that melting of different
quarkonium states does not lead to significant change in the Euclidean time
correlators. The large change of the quarkonium correlators above deconfinement 
observed in the scalar and axial-vector channels appears to be due to the zero
mode contribution.
\end{abstract}


\section{Introduction}
\label{intro}

It was argued long ago that melting of quarkonia
above the deconfinement transition can serve as a 
signature of quark-gluon plasma formation in heavy
ion collisions \cite{MS86}. The basic idea behind this
proposal was that due to color screening the potential 
between a quark and anti-quark will not provide sufficient
binding at high temperature. This problem can be formulated
more rigorously in terms of quarkonium spectral functions, which can be, 
in principle, extracted from Euclidean-time meson correlation functions calculated on the lattice.
Attempts doing this based on the Maximum Entropy Method 
(MEM) have been discussed over the last few
years. The initial interpretation of data led to the conclusion that the $1S$ charmonium states
survive in the deconfined medium up to temperatures of about
$1.6T_c$, with $T_c$ being the transition temperature 
\cite{umeda02,asakawa04,datta04,swan}. Recent analysis, however,
has shown that, although  MEM can be used to extract reliably  
quarkonium spectral functions at zero temperature, at finite 
temperature it has severe limitations \cite{jako07}.

At zero temperature quarkonium spectrum is well described in non-relativistic 
potential models. 
Since the seminal paper by Matsui and Satz the problem of charmonium
dissolution has been studied in potential models 
\cite{karsch88,ropke88,hashimoto88,digal01,wong04,mocsy05,mocsy06,rapp06,alberico07,wong06}. 
While the early  studies used phenomenological potential, more recent studies rely on 
static quark anti-quark free energy calculated on the lattice. In fact significant
progress has been made in understanding the in-medium modification of inter-quark forces via
lattice calculations of the free energy of static quark anti-quark pair. Calculations have been done in pure gluodynamics, 3-flavor and 2-flavor  
QCD \cite{okacz,petrov04,okacz05}, and preliminary results are also available in 
the physically relevant case of one heavy strange quark 
and two light quarks  \cite{kostya,olaf07} (the light quark masses correspond to pion mass of about $220$~MeV).

Recently attempts to calculate
quarkonium properties at finite temperature using resummed perturbation theory have been made
\cite{laine06,laine07}. Resummed perturbation theory appears to be successful in calculations of bulk
thermodynamics properties \cite{scpt97,braaten,blaizot}.

Since the lattice calculations of spectral functions have severe limitations, 
in \cite{mocsy05,mocsy06} it has been pointed out, that comparison between the lattice data and potential models
should be done in terms
of the Euclidean time correlators, for which the numerical
results are much more reliable. Recent studies 
following this line have also been presented in Refs. \cite{rapp06,alberico07,{wong06}}. 
In this contribution we discuss the calculation of quarkonium spectral functions in a potential
model, which uses the lattice data of the free energy of a static quark anti-quark pair.
Since reliable calculations of the quarkonium correlators are available only in quenched approximation
we consider QCD with only heavy quarks. Further details about this approach can be found
in Ref. \cite{mocsy07}.

\section{Charmonium Spectral Functions in Potential Model}

For heavy quarks the spectral function can be related to the non-relativistic Green's function 
\begin{eqnarray}
&
\displaystyle
\sigma(\omega)=K \frac{6}{\pi} {\rm Im} G^{nr}(\vec{r},\vec{r'},E)|_{\vec{r}=\vec{
r'}=0}\, ,\\[2mm]
&
\displaystyle
\sigma(\omega)=K \frac{6}{\pi}\frac{1}{m^2} {\rm Im} \vec{\nabla} 
\cdot \vec{\nabla'} G^{nr}(\vec{r},\vec{r'},E)|_{\vec{r}=\vec{r'}=0}\, ,
\label{green_sc} 
\end{eqnarray}
for $S$-wave, and $P$-wave charmonia, respectively. Here $E=\omega-2 m~$.
At leading order $K=1$. Relativistic and higher order perturbative corrections
will lead to a value different from unity \cite{mocsy07}.
The non-relativistic Green's function satisfies the Schr\"odinger equation
\begin{equation}
\left [ -\frac{1}{m  } \vec{\nabla}^2+V(r)-E \right ] G^{nr}(\vec{r},\vec{r'},E) = \delta^3(r-r')\, .
\label{schroedinger}
\end{equation}
The numerical method for solving this equation is presented in \cite{mocsy07}.
At zero temperature we use the Cornell potential $V(r)=-\alpha/r+\sigma r$ with parameters 
motivated by lattice results on static potential : $\alpha=\pi/12$ and 
$\sigma=(1.65-\pi/12)r_0^{-2}$.
In the actual calculations we use a potential which is screened beyond some distance $r>r_{med}$
with screening length $\mu$ to mimic many body effects at large energies 
 (see Ref. \cite{mocsy07} for further details).   
At finite temperature we use a potential motivated by lattice results on the
singlet free energy of a static quark anti-quark pair and which is defined in section 
IV of Ref. \cite{mocsy07}.
At large energies, away from the threshold, the non-relativistic treatment is  
not applicable. 
The spectral function in this domain, however,  can be calculated using perturbation theory. 
We smoothly match the non-relativistic calculation of the 
spectral function to the relativistic perturbative result \cite{mocsy07}.  
Euclidean time correlators $G(\tau,T)$ at some temperature $T$ can 
be calculated from the spectral functions
using the integral representation
\begin{equation}
G(\tau,T)=\int_0^{\infty} d \omega \sigma(\omega,T) K(\omega,\tau,T)\, .
\label{spect_rep}
\end{equation}
Here the integration kernel is 
\begin{equation}
\displaystyle
K(\omega,\tau,T)=\frac{\cosh \omega (\tau-1/(2T))}{\sinh \left( \omega/(2T) \right)}\, .
\label{kernel_T}
\end{equation}

\section{Correlators at Zero Temperature}
\begin{figure}[htb]
\includegraphics[width=9.1cm]{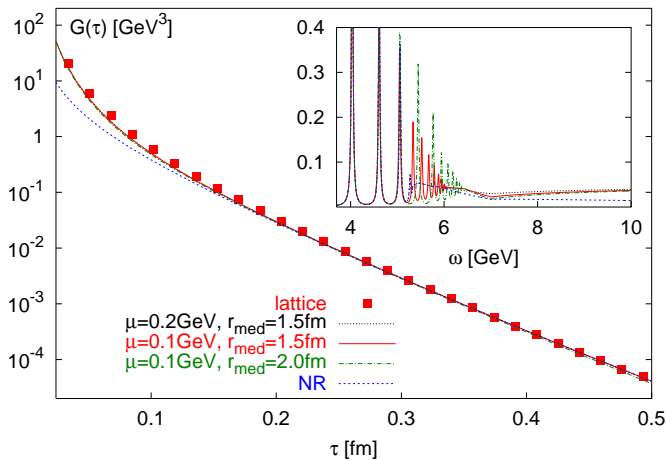}
\caption{The pseudo-scalar charmonium correlator calculated in our
  model and compared to the  lattice data of Ref. \cite{datta04}. In the
  inset, the corresponding spectral functions
  $\sigma(\omega)/\omega^2$ are shown.}
\label{fig:pscorr}
\end{figure}

In this section we discuss the comparison of the model calculations with zero temperature
lattice data from isotropic lattices \cite{datta04}. The lattice spacing has been
fixed using the Sommer-scale $r_0=0.5$fm. Its value is slightly larger than the one used in Ref. 
\cite{datta04}, since there the string tension of $\sqrt{\sigma}=420~$MeV has been 
used to set the scale. Calculations have
been done at the charm quark mass which corresponds to an $\eta_c$ mass of about $4~$GeV.
The renormalization constants of the lattice operators have been calculated in 1-loop tadpole
improved perturbation theory (see Ref. \cite{datta04} for further details). 
The $K$ factors in Eq. (\ref{green_sc}) have been chosen such that at large distances the
correlators calculated in potential models agree with the lattice results.  
In Fig. \ref{fig:pscorr} we show the pseudo-scalar correlator calculated on the lattice and in the
potential model for several screening parameters, together with the corresponding spectral functions.
As one can see from the figure the choice
of the ad-hoc screening parameters have almost no effect on the Euclidean correlator.
We see a reasonably good agreement between the lattice data and potential model calculations.
Also shown in the figure is the correlator corresponding only to the non-relativistic spectral
function. At small Euclidean times this falls below the lattice data by more than an order of magnitude. Thus correlators calculated on the lattice are sensitive to the relativistic
continuum part of the spectral functions. Similar analysis of the correlators have been done in the vector,
axial vector and scalar channels. 

\section{Temperature-dependence of Quarkonium Correlators}

\begin{figure}[htb]
\includegraphics[width=7.7cm]{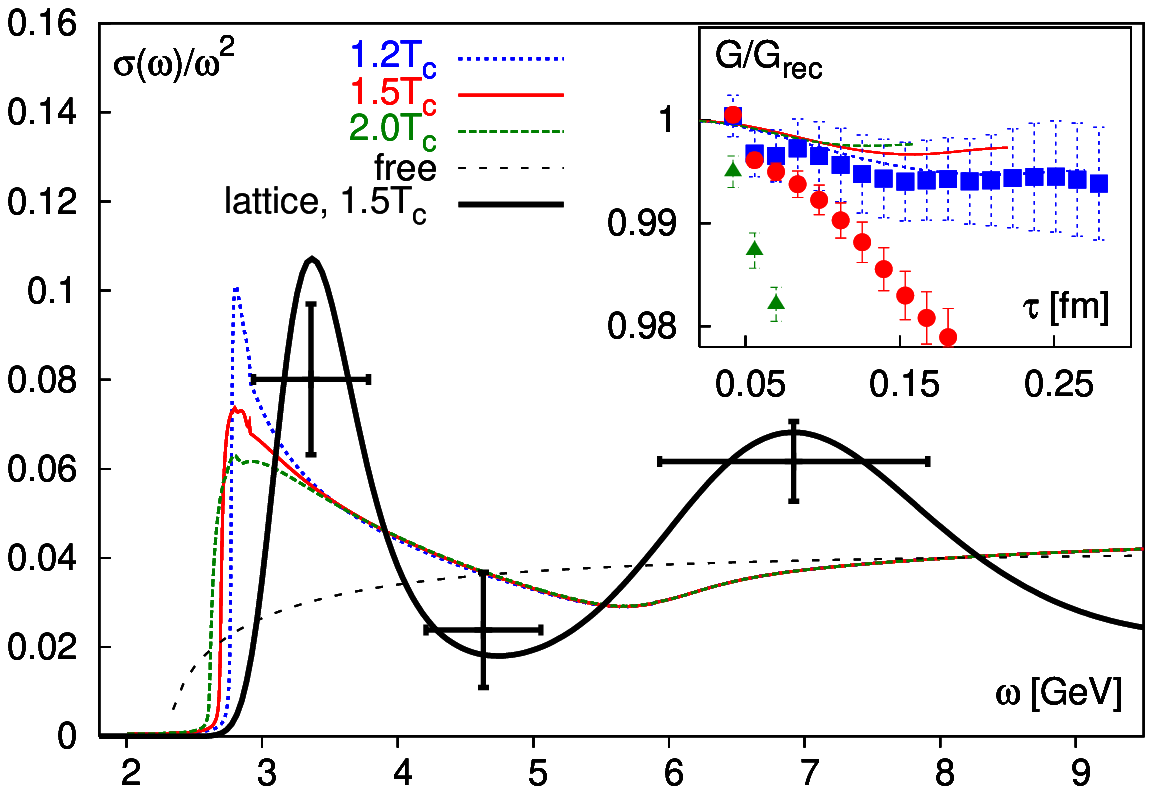}
\includegraphics[width=7.7cm]{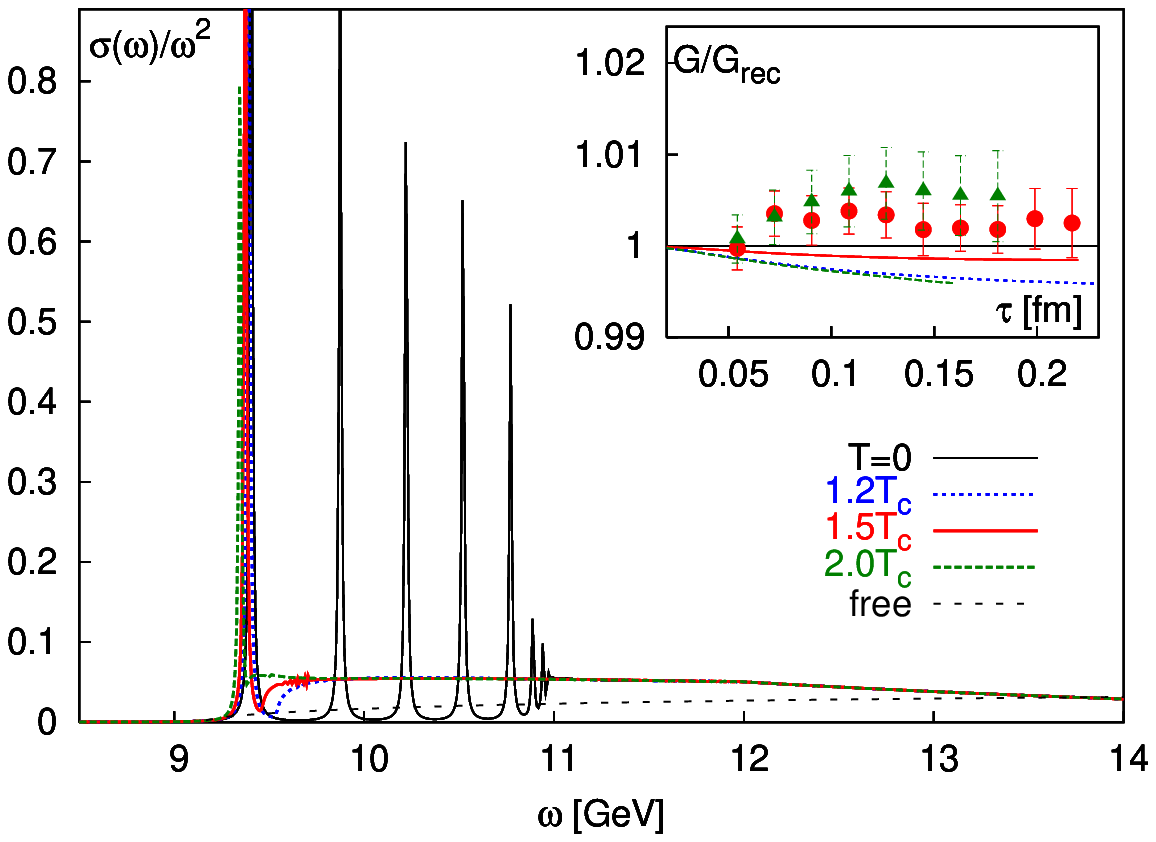}
\caption{The charmonium (left) and bottomonium (right) spectral
  functions at different temperatures. 
  For charmonium we also show the spectral functions
  from lattice QCD obtained from the MEM at $1.5T_c$. The error-bars
  on the lattice spectral function correspond to the statistical error
  of the spectral function integrated in the $\omega$-interval
  corresponding to the horizontal error-bars.  The insets show the
  corresponding ratio $G/G_{rec}$ together with the results from
  anisotropic lattice calculations \cite{jako07}.  For charmonium,
  lattice calculations of $G/G_{rec}$ are shown for $T=1.2T_c$
  (squares), $1.5T_c$ (circles), and $2.0T_c$ (triangles). For
  bottomonium lattice data are shown for $T=1.5T_c$ (circles) and
  $1.8T_c$ (triangles).  }
\label{fig:res_v_ps}
\end{figure}

In this section we study the temperature-dependence of quarkonium
spectral functions and correlators. Since the correlators depend on
the temperature through the integration kernel
and the spectral functions, it is customary to
study the temperature dependence of the correlators in terms of 
the ratio $G(\tau,T)/G_{rec}(\tau,T)$, where
\begin{equation}                                                                                             
G_{rec}(\tau,T)=\int_0^{\infty} d \omega \sigma(\omega,T=0) K(\omega,\tau,T)\, .        
\label{grec} 
\end{equation}
This way the trivial temperature dependence due to the integration kernel is taken care of.
Also many uncertainties of the lattice calculations cancel out in this ratio. 
The finite temperature spectral functions are shown in Fig. \ref{fig:res_v_ps}
for the pseudo-scalar channel. The 1S charmonium state is melted at $1.2T_c$.
We see, however, a large enhancement of the spectral functions near the threshold.
Note, that the height of the spectral functions near the threshold is comparable to the height
of the bump in the spectral function calculated from MEM \cite{jako07}. It is therefore possible
that the bump of the spectral functions calculated from lattice correlators using MEM actually
corresponds to a threshold enhancement, and was mistakenly interpreted as the $1S$ state.
In the case of bottomonium all states, except the $1S$ state, are dissolved above the
deconfinement transition. The $1S$ state can survive as a resonance until temperatures of about $2T_c$.
\begin{table}
\begin{tabular}{|ccccc|}
\hline
$T$             &    0           &    $1.2T_c$    &     $1.5T_c$     &        $2.0T_c$  \\
\hline             
$s_0$         &  10.975     &    9.541         &      9.462           &         9.384 \\
$M(1S)$     &  9.405       &   9.390          &      9.374           &         9.343 \\        
$E_{bind} $ &  1.570       &  0.151          &       0.088           &        0.041\\
\hline
\end{tabular}
\caption{The mass and the binding energy of the $1S$ bottomonium state at different temperatures, and the continuum threshold.}
\label{tab}
\end{table}
Note, however, that the binding energy of the $1S$ bottomonium is significantly reduced due to color
screening, as shown in Table \ref{tab}.
The binding energy is defined as the distance between the continuum threshold $s_0$ and the bound state peak.
 Due to
the small binding energy the 1S state will acquire a sizable thermal width, and may not show up
as the resonance in the corresponding spectral function. Therefore, the actual dissolution temperature
of the $1S$ bottomonium will be smaller then the one estimated based on the simple potential
model calculations which do not include the effect of the thermal width \cite{mocsy07a}. 
In the insets of Fig. \ref{fig:res_v_ps} we also show the corresponding ratio $G/G_{rec}$.
The large changes in the spectral functions are not visible in the correlation functions. These 
agree quite well with the lattice results. This is because even in the absence of bound states the
spectral function is significantly larger than the spectral function corresponding to a freely propagating
quark anti-quark pair.
\begin{figure}[htb]
\includegraphics[width=7.6cm]{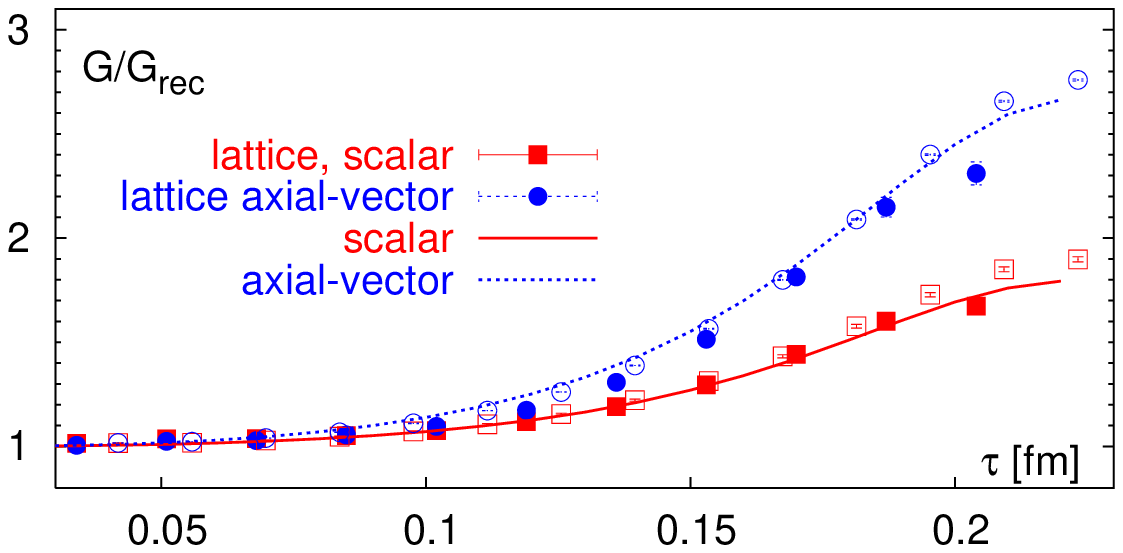}
\includegraphics[width=7.6cm]{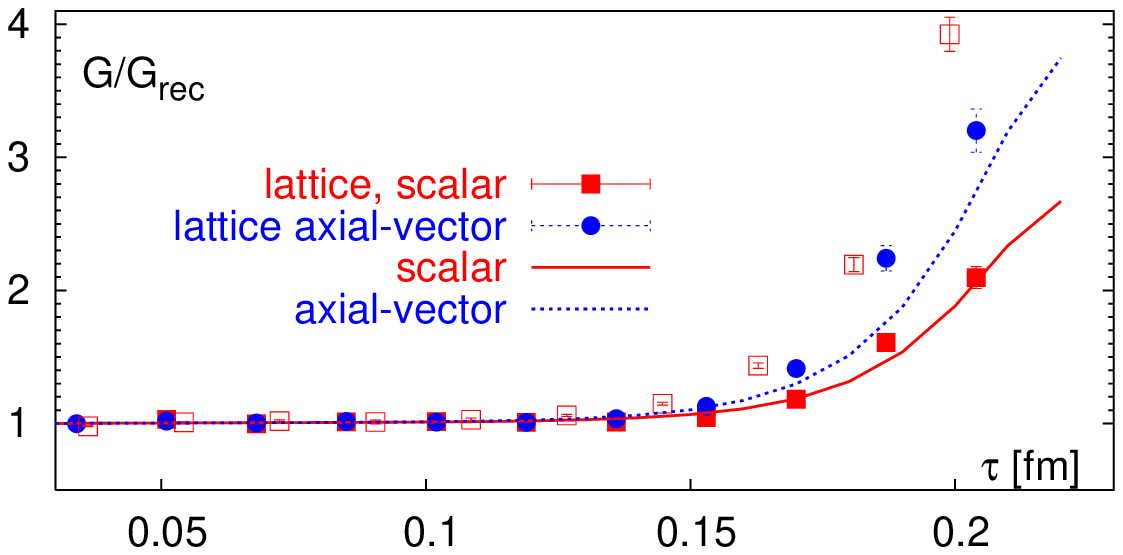}
\caption{The ratio $G/G_{rec}$ in the scalar and axial-vector channel
  at $T=1.5T_c$ for charmonium (left) and bottomonium (right).
  Lattice calculation on isotropic lattices
  \cite{datta04,datta_panic05} are shown as filled symbols. Open
  symbols refer to results from anisotropic lattice calculations of
  Ref. \cite{jako07}.}
\label{fig:resP}
\end{figure}
In the vector, scalar and axial-vector channels significant temperature dependence has
been found \cite{datta04,jako07,datta_panic05}. It has been shown that this is due to
the zero mode contribution, i.e. due to the $\omega \delta(\omega)$-like contribution to the
quarkonium spectral functions \cite{umeda07}. In the vector channel the zero mode contribution
corresponds to the heavy quark transport \cite{derek}.
The zero mode contribution can be estimated in the free case. It can also be shown that the zero mode contribution
is absent
in the pseudo-scalar channel.
If we add the free theory result for the zero mode contribution to the spectral function calculated in the potential model we can reproduce the
temperature dependence of the scalar and axial-vector correlator both for charmonium and bottomonium.
This is demonstrated by Fig. \ref{fig:resP} where $G/G_{rec}$ for the scalar and axial-vector channels is shown.

\section{Conclusions}
We discussed the calculations of quarkonium spectral functions and the corresponding Euclidean
time correlators in a potential model. We have found that all quarkonium states, except the $1S$ 
bottomonium state, dissolve at temperatures smaller than about $1.2T_c$.  This, however, does not lead
to significant change of the correlators. Zero mode contribution on the other hand could give a large
change in the correlators above the deconfinement transition. We have found that the spectral  
functions calculated in our model can explain quite well the temperature dependence of the quarkonium
correlators obtained in lattice QCD.

\section*{Acknowledgments}

This work has been supported by U.S. Department of Energy under Contract No. DE-AC02-98CH10886.

\end{document}